%% file: main.tex
\RequirePackage{fix-cm}

\documentclass[twocolumn]{svjour3}          % twocolumn
\smartqed  % flush right qed marks, e.g. at end of proof
\usepackage{svg}
\usepackage{hyperref}       % hyperlinks
\usepackage{graphicx}
\usepackage{amsmath}
\usepackage{subcaption}
\usepackage{color}
\usepackage[numbers]{natbib}
\usepackage{glossaries}

\def\centerarc[#1](#2)(#3:#4:#5)% Syntax: [draw options] (center) (initial angle:final angle:radius)
{ \draw[#1] ($(#2)+({#5*cos(#3)},{#5*sin(#3)})$) arc (#3:#4:#5); }

\usepackage{csvsimple}
\usepackage{changepage}
\usepackage[tbtags]{mathtools}
\usepackage{siunitx}

\usepackage{csquotes}

\usepackage{multirow}
\usepackage{tabularx}
\usepackage{booktabs}
\usepackage{array}
\usepackage{longtable}
\newcolumntype{L}[1]{>{\raggedright\let\newline\\\arraybackslash\hspace{0pt}}m{#1}}
\newcolumntype{C}[1]{>{\centering\let\newline\\\arraybackslash\hspace{0pt}}m{#1}}
\newcolumntype{R}[1]{>{\raggedleft\let\newline\\\arraybackslash\hspace{0pt}}m{#1}}

\DeclareSIUnit\pixel{px}

\definecolor{someprettyred}{rgb}{0.85, 0.0, 0.1}

\newacronym{ct}{CT}{computed tomography}
\newacronym{3d}{3D}{three-dimensional}
\newacronym{2d}{2D}{two-dimensional}

\usepackage{pgfplots}
\usepackage{subcaption}
\usepackage{tikz}
\usepgfplotslibrary{external} 
\usetikzlibrary{arrows}
\pgfplotsset{compat=newest}
\usepgfplotslibrary{fillbetween}
\usetikzlibrary{math}
\usetikzlibrary{calc}
\usetikzlibrary{patterns}
\def\centerarc[#1](#2)(#3:#4:#5)% Syntax: [draw options] (center) (initial angle:final angle:radius)
{ \draw[#1] ($(#2)+({#5*cos(#3)},{#5*sin(#3)})$) arc (#3:#4:#5); }
\usetikzlibrary{shapes.geometric}
\usetikzlibrary{positioning}
\usetikzlibrary{shapes}
\usetikzlibrary{intersections}
\usetikzlibrary{arrows.meta}
\tikzset{font=\small}
    \pgfdeclarepatternformonly{south west lines}{\pgfqpoint{-0pt}{-0pt}}{\pgfqpoint{3pt}{3pt}}{\pgfqpoint{3pt}{3pt}}{
        \pgfsetlinewidth{0.4pt}
        \pgfpathmoveto{\pgfqpoint{0pt}{0pt}}
        \pgfpathlineto{\pgfqpoint{3pt}{3pt}}
        \pgfpathmoveto{\pgfqpoint{2.8pt}{-.2pt}}
        \pgfpathlineto{\pgfqpoint{3.2pt}{.2pt}}
        \pgfpathmoveto{\pgfqpoint{-.2pt}{2.8pt}}
        \pgfpathlineto{\pgfqpoint{.2pt}{3.2pt}}
        \pgfusepath{stroke}}

\input{color_define.tex}

\title{Enhanced Interlocking in Granular Jamming Grippers through Hard and Soft Particle Mixtures}
\subtitle{}

\author{Angel Santarossa  \and
        Thorsten P\"oschel
}

\institute{Angel Santarossa \and Thorsten P\"oschel \at
Institute for Multiscale Simulations\\
Friedrich-Alexander-Universit\"at Erlangen-N\"urnberg\\
Cauerstra\ss{}e 3, 91058 Erlangen\\
Germany\\
\email{thorsten.poeschel@fau.de}           %  \\
}

\date{Received: date / Accepted: date}
\begin{document}
\sloppy

\maketitle

\begin{abstract}
%Granular grippers are soft robotic effectors that exploit the effect of granular jamming to grip objects. Their gripping strength arises from friction, geometrical constraints, and suction effects. In this work, we experimentally study the effect of particle stiffness on the interlocking mechanism in granular grippers. Using X-ray imaging, we show that in contrast to hard particles, a gripper filled with soft particles squeezes the object after jamming is induced, wrapping around its protrusions. This results in significant holding forces due to interlocking when the object is pulled. Additionally, we show that a gripper filled with a small fraction of hard particles within a packing of soft particles increases considerably further the maximum holding force achieved by the gripper through interlocking.  

We investigate the influence of particle stiffness on the grasping performance of granular grippers, a class of soft robotic effectors that utilize granular jamming for object manipulation. Through experimental analyses and X-ray imaging, we show that grippers with soft particles exhibit improved wrapping of the object after jamming, in contrast to grippers with rigid particles. This results in significantly increased holding force through the interlocking. The addition of a small proportion of rigid particles into a predominantly soft particle mixture maintains the improved wrapping but also significantly increases the maximum holding force. These results suggest a tunable approach to optimizing the design of granular grippers for improved performance in soft robotics applications.

\end{abstract}

\section{Introduction}
\label{sec:intro}

Granular grippers are soft robotic effectors that exploit granular jamming to grip objects \cite{Brown2010universal}. These grippers are highly adaptable and can easily manipulate objects of diverse geometry and surface properties without the need for reconfiguration between gripping cycles\cite{amend2012positive, amend2016soft, miettinen2019granular, licht2018partially, kapadia2012design, kremer2023trigger, fitzgerald2020review}. 
The typical design of a granular gripper consists of a granulate that is held in place by a flexible, hermetically sealed membrane.
%The typical design of a granular gripper features granulate enclosed within a flexible, hermetically sealed membrane. 
%In this state, the granular material is plastically deformable. 
In this state, the granular material easily flows when deformed, similar to a fluid. When pressed against an object, the gripper deforms and adapts to the object's contour. 
When the air is pumped out of the gripper, the membrane contracts and compresses the particles. This causes the granulate to jam and assume a mechanically robust, solid-like state characterized by a pronounced modulus of elasticity. The stiffness of the system increases significantly in the jammed state \cite{liu2001jamming, Jaeger2015towards}. As the material jams, the gripper exerts force on the target object, that allows it to be gripped and held effectively. To release the object, the vacuum is broken, causing the membrane to relax and the granules to return to its deformable state.

The holding force that a granular gripper generates when engaging with an object is influenced by the mechanical characteristics of the target object. It emerges from the interplay of three distinct mechanisms\cite{Brown2010universal}: static friction, suction, and geometrical interlocking. Friction arises from tangential stress at the contact between the object and the gripper membrane \cite{gomez2021effect, gotz2022soft}. Suction is activated when the membrane seals the object's surface airtight \cite{santarossa2023effect}. When the gripper is retracted, the pressure within the sealed gap decreases, resulting in an attractive force on the object. Interlocking occurs when the membrane encloses the object (or any of its protrusions), creating geometric constraints with the object as the granulate hardens, effectively securing the grip \cite{kapadia2012design}.

In granular grippers, the hoisting forces due to interlocking and suction are considerably larger than those produced through friction, with interlocking being the most efficient mechanism \cite{Brown2010universal, lichtUniversalJammingGrippers2016}. Thus, in cases lacking suction and interlocking, the grippers holding capability relies solely on friction, resulting in notably reduced performance. Strengthening the locking mechanism is, therefore, highly desirable. However, the contribution of interlock to the hoisting force depends significantly on the object's shape and the mechanical characteristics of the granulate. The granular material needs to flow and adapt to the object's contours, thereby establishing geometric constraints that improve the grip as jamming occurs.

To disengage interlocking, the gripper must flexibly relax, allowing the gripped object to be released smoothly. It was suggested that the peak hoisting force generated by interlocking is directly related to the strength of the granular material in its jammed state. This strength is specifically related to the maximum stress the jammed granular material can sustain under bending and stretching \cite{Brown2010universal}. In this sense, it was shown \cite{gotz2023granular} that the resistance to bending of a simply supported beam made of jammed granular matter encased in a membrane increases with the particle stiffness. Consequently, the use of rigid particles is anticipated to significantly enhance the interlocking capability of granular grippers. 

Brown et al. \cite{Brown2010universal} measured the hoisting force of a gripper filled with glass beads when gripping a sphere. They found that the contribution of interlocking to the hoisting force increases with the extent to which the gripper envelops the object, quantified by the contact angle. In their experiments, geometrical interlocking was achieved by manually molding the gripper in the unjammed state around the object. Such manual manipulation of the gripper is impractical for real-world applications, like handling objects in factory automation.

In prior research \cite{gotz2022soft}, we demonstrated that, unlike rigid particles, incorporating soft particles in granular grippers results in a distinct squeezing effect. Specifically, the gripper exerts significant compression on the object following the induction of jamming. This effect causes substantial normal and, consequently, frictional forces, contributing to large hoisting capability. However, the influence of particle elasticity on the interlocking mechanism remains unexplored.

In the present work, we show that soft particles, as opposed to rigid ones, significantly enhance the hoisting forces of granular grippers through interlocking. Utilizing X-ray computed tomography (CT), we reveal that grippers filled with soft particles undergo a notable reduction in volume compared to those with rigid particles. This reduction allows the gripper to more effectively envelop object protrusions or convex surfaces. Introducing a small proportion of rigid particles into a predominantly soft particle-filled gripper substantially enhances interlocking. X-ray CT scans reveal that the gripper's volume reduction is caused by soft particles, which, in turn, facilitates the displacement of the more rigid particles around the object's protrusions, largely increasing the force required to break the interlock.

\section{Materials and methods}\label{secmethods}

\subsection{Experimental setup and measurement procedure}\label{sec1_2_methods}

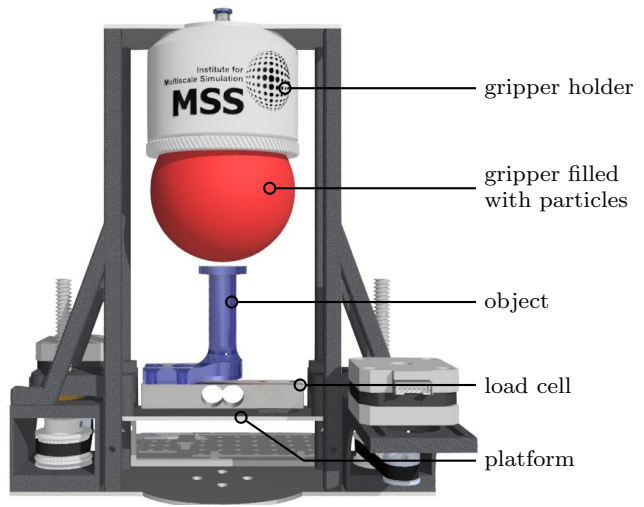
\begin{figure}[htb]
\centering
    \input{setup}
    \caption{\label{fig:expSetup}Experimental setup. The gripper, consisting of a granulate contained within an elastic membrane, grips an object positioned on a platform below the gripper. The platform's vertical position can be adjusted. A load cell captures the force exerted on the object.}
\end{figure}

Figure \ref{fig:expSetup} sketches the experimental setup: The granular gripper consists of a spherical elastic, air-tight bag with diameter $(70.0 \pm 0.5) \text{mm}$, filled with granulate. Two types of beads are used for the experiments: glass beads of diameter $(4.0 \pm 0.3) \text{mm}$ and expanded polystyrene (EPS) beads of diameter $(4.2 \pm 0.5) \text{mm}$. The gripper bag is filled with $(200 \pm 1) \text{cm}^{3}$ of grains and is attached to a holder. A flange on the holder's top allows air pressure or vacuum to be applied within the gripper. Positioned beneath the gripper, the target object rests on a platform that can move vertically. The base's vertical position can be adjusted via two stepper motors driven by a microcontroller. A load cell, placed between the object and the platform, measures the force exerted on the target object.

We examine the gripping process involving a cylinder-symmetric object of a T-shaped cross-section, as depicted in Fig. \ref{fig:expSetup}. The specific shape of the object enables the gripper to envelop and interlock with it. The object consists of two cylindrical sections: a lower cylinder of diameter $(14.0 \pm 0.1) \text{mm}$ and height $(44.0 \pm 0.1) \text{mm}$, and an upper flat cylinder of diameter $(23.9 \pm 0.1) \text{mm}$ and height $(3.5 \pm 0.1) \text{mm}$. To inhibit the suction effect, the object's surface was intentionally roughened, thus, no air tight cavities could develop.

We studied the gripping performance using granulates of different compositions, namely, (a) glass beads, (b) EPS particles, and (c) different binary mixtures of both (fraction of glass beads: 
%\textcolor{red}{
0.05, 0.1, 0,15, 0.20, 0.25, 0.5, and 0.75). 
%For the binary mixtures, the gripper is first filled with glass and then with EPS beads.}   

Before each measurement, the gripper is pressurized to fluidize the enclosed granulate, effectively resetting any memory of previous gripping cycles. Each measurement comprises the following steps:
\begin{itemize}
    \item[(i)]  Initially, the platform is raised, pressing the object against the gripper, which molds to the object's shape. Once a predetermined indentation depth is achieved, the upward motion of the platform halts.
    \item [(ii)] The granulate is then given a brief period to settle, during which the particles adjust and realign. 
    \item [(iii)] The air is evacuated from the gripper, creating a pressure differential between the gripper's interior and the external atmosphere, $p_\text{vac}\approx 90 \text{kPa}$. This pressure difference causes the membrane to tighten and compress the granulate, triggering jamming. The specific pressure value chosen corresponds to the maximum attainable by the vacuum pump. It was shown that large differential pressure corresponds to large values of the hoisting force of granular grippers \cite{lichtUniversalJammingGrippers2016}.
    \item [(iv)] Finally, the platform is lowered until the object is no longer in contact with the gripper. The force applied to the object is continuously recorded throughout the gripping process.  In particular, we consider the peak hoisting force attained by the gripper.
    \end{itemize}

\subsection{X-ray computed tomography}
The hoisting force generated by the gripper reflects the macroscopic response of the gripper due to the internal motion of the particles during the gripping process. To relate this macroscopic response to the rearrangement of the particles in the course of the gripping process, the granular packing post-jamming is examined using X-ray computed tomography (CT).

CT scans are conducted with a high-capacity laboratory X-ray scanner, accommodating the full experimental setup. The parameters chosen for the CT scans are detailed in Table \ref{tab:parameters_ct1}. To mitigate beam hardening effects, a copper plate of \SI{1.5}{\mm} thickness is used as a filter \cite{baur_correction_2019}. The granular packing's three-dimensional (3D) structure is then reconstructed using the software XrayOffice (v2.0).
\begin{table}[htb]
\centering
\caption{Parameters used for X-ray CT\label{tab:parameters_ct1}}
\begin{tabular}{lrl}
 parameter & value\\\hline
 source voltage & 140 & kV\\
 target current & 320 & $\mu$A\\
 projections per scan &  1600 \\
 measurements per projection & 10\\
 exposure time & 150& ms\\
 resolution & 85.4 & $\mu$m$^{-1}$ \\
\end{tabular}
\end{table}

\section{Results and discussions}\label{sec2_results}
\begin{figure}[ht]
\centering
    \input{holding-force-EPS}
    \caption{Maximum holding force achieved when gripping a screw-shaped object using a gripper filled with glass and EPS beads. Each result is averaged over six independent measurements. The error bars indicate the standard deviation of the mean.
\label{fig:HF-softP-interlocking}}
\end{figure}
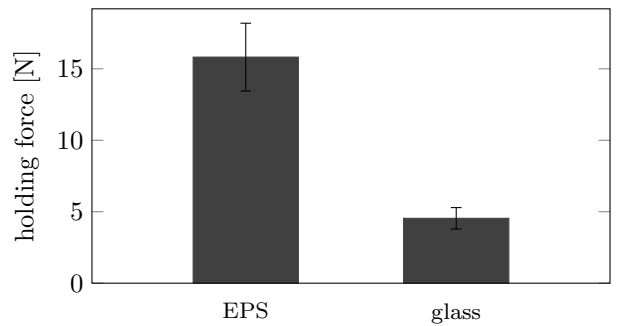

We consider the impact of particle stiffness on the hoisting force of a granular gripper. Figure \ref{fig:HF-softP-interlocking} shows the peak hoisting forces of a gripper filled with either glass or EPS (Expanded Polystyrene) beads. The maximum holding force attained with EPS is considerably larger ($\sim \text{3.5}$ times) than for glass beads, highlighting the enhanced role of interlocking when using EPS particles.
At first glance, the result appears counterintuitive. If geometrical constraints are formed between the gripper and the target object in both cases -- glass beads and EPS beads -- one would expect a stronger hoisting force for more rigid particles. This expectation follows from the logic that rigid particles confer greater stiffness to the gripper, which requires greater forces to deform the gripper sufficiently for object release \cite{Brown2010universal}. 
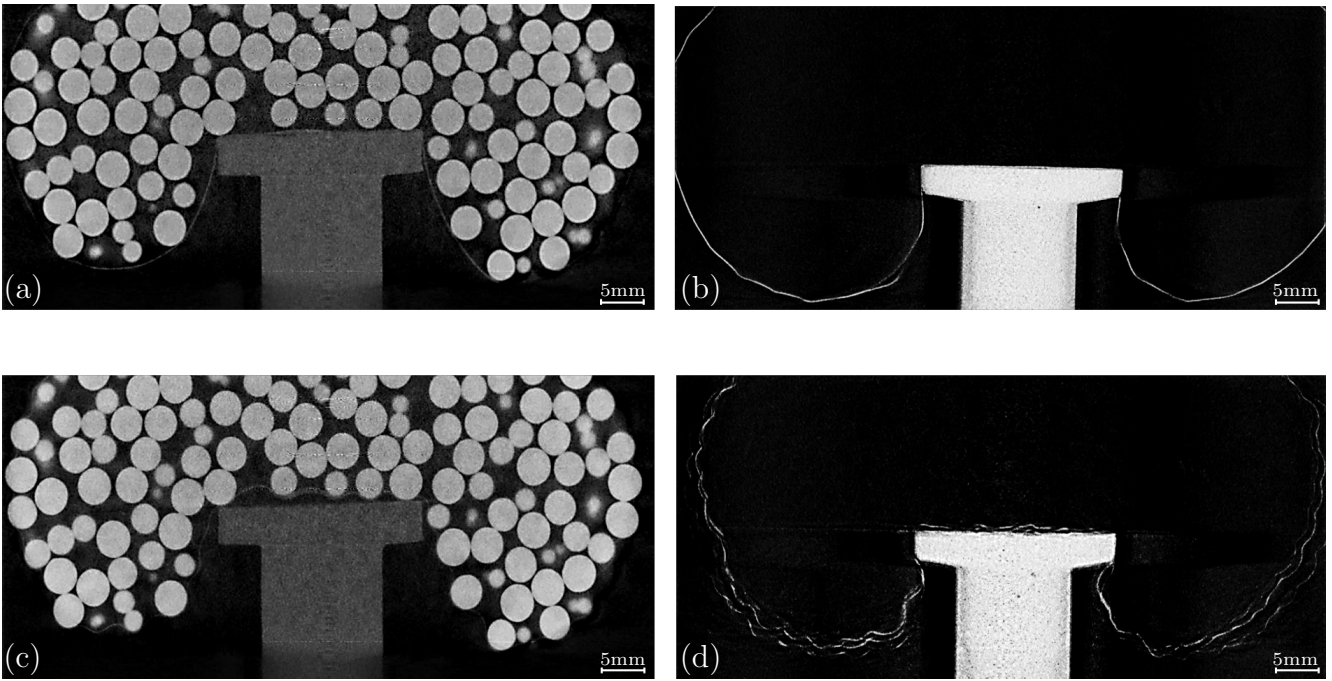
\begin{figure*}[t!]
\centering
    \input{softP-xray-interlocking}
    \caption{X-ray tomogram slices of the gripper filled with glass beads (a,c) and EPS beads (b,d) before applying vacuum (a,b) and after (c,d). EPS particles are not visible due to their low X-ray absorbance. The white contour represents the gripper membrane. 
\label{fig:x-ray-softP-interlocking}}
\end{figure*}

To elucidate the enhanced hoisting force observed with soft particles, we examine the granular packing within the gripper at various stages of the gripping process. Tomograms are captured at two critical junctures: (1) after the gripper has molded around the object before evacuating the gripper and (2) after the evacuation and subsequent jamming. Figure \ref{fig:x-ray-softP-interlocking} shows vertical sections through the granular packing inside the gripper for both glass beads (Fig.\ref{fig:x-ray-softP-interlocking}(a,c)) and EPS beads (Fig.\ref{fig:x-ray-softP-interlocking}(b,d)), before and after the air is evacuated. For EPS particles, the volume of the packing reduces drastically ($\approx 28 \%$), which enables the gripper to closely encase the object's protrusions, facilitating geometrical interlocking. In contrast, for glass beads, we see negligible volume change after evacuation (less than 1\%). Consequently, in the process of jamming, the gripper hardly conforms around the object's protrusions, impeding interlocking. 

\begin{figure}[ht]
\centering
    \input{holding-force-mixtures2}
    \caption{Maximum holding force of a gripper filled with mixtures of glass and EPS beads at different volume fractions (black bars). Blue bars show the results for pure EPS and pure glass beads. Averages are due to six independent measurements, and error bars show the standard deviation of the mean. 
\label{fig:HF-mixtures-interlocking}}
\end{figure}
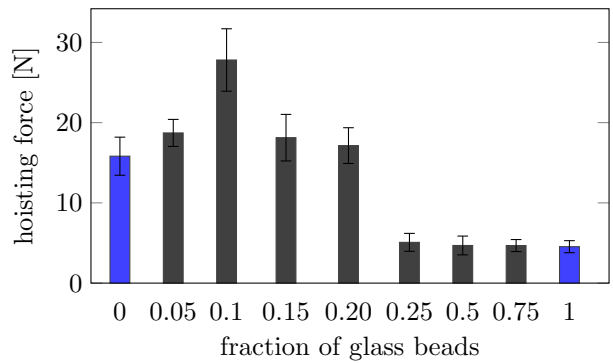

The hoisting force of a gripper filled with a mixture of hard gall and soft EPS beads can substantially exceed the values of pure glass and pure EPS beads. Figure \ref{fig:HF-mixtures-interlocking} shows the maximum holding forces achieved by a gripper filled with mixtures of glass and EPS beads at different fractions. 
The peak force, approximately 28 N, is attained at a 0.1 volume fraction of glass beads, highlighting the substantial role of interlocking to the hoisting force. Remarkably, the peak force for a mixture exceeds the values for pure granulate considerably, about six times for pure glass beads (Fig. \ref{fig:HF-softP-interlocking}) and two times for pure EPS beads (Fig. \ref{fig:HF-softP-interlocking}). 
For glass concentration $\geq 0.25$ (Fig. \ref{fig:HF-mixtures-interlocking}), the hoisting force is $\approx 5\,\text{N}$, independently of the fraction of glass beads. From these results, we conclude that soft particles support the interlocking, while hard particles provide the final force.  

To explain the increase of the hoisting force, we use X-ray imaging. 
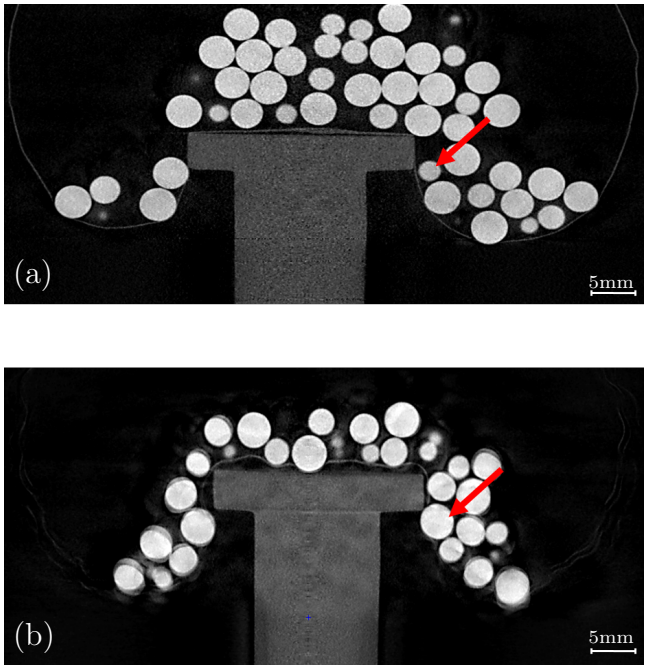
\begin{figure}[ht]
\centering
    \input{mixture-x-ray-interlocking}
    \caption{X-ray tomograms before evacuation (a) and after (b). The gripper contains a mixture of glass beads and 10\% EPS beads. The EPS particles are not visible due to their low X-ray absorbance. The white contour shows the gripper membrane. The red arrows highlight a particle that contributes to interlocking once jamming is induced.
\label{fig:x-ray-mixture-interlocking}}
\end{figure}
Figure \ref{fig:x-ray-mixture-interlocking} shows tomogram slices through the gripper filled with a mixture of glass and EPS beads with a glass fraction  of 0.1 before and after the evacuation of the air. After evacuation, the gripper's volume reduces considerably, allowing it to embrace the object's protrusions, facilitating interlocking. Since glass beads are closer to the object, the gripper shrinkage causes the displacement of the more rigid particles around the object's edges. Therefore, geometrical interlocking is predominantly achieved by rigid particles, which significantly increases the force needed to break the interlock.

%Mixing between soft and hard particles can occur after numerous gripping cycles. In such a case, interlocking is achieved through a combination of soft and hard particles, decreasing the maximum holding force achieved. For practical applications, a gripper bag with a compartmentalized structure could be produced to prevent the mixing: A lower compartment can allocate the hard particles, while an upper compartment holds the soft beads.

\section{Summary and outlook}\label{sec3}

We studied the hoisting force of a granular gripper and find that it strongly depends on the composition of the granulate. For a mixture of soft and hard particles, the hoisting force exceeds the values obtained for pure granulates, hard or soft. The maximum force is achieved for a 10\% volume fraction of hard beads. 

This increase is explained by the interlocking mechanism in granular grippers, which has previously been identified as the main contribution to the hoisting force \cite{gotz2022soft}. While soft particles do not contribute much to the strength of the jammed granulate as a meta-material \cite{gotz2023granular,PhysRevResearch.6.013061}, they enable the granulate to shrink in volume during the evacuation of the gripper. This shrinkage corresponds to a non-negligible motion of the particles, which in turn allows the granular gripper to better embrace the object. Both mechanisms, shrinkage and embracement by the dislocation of particles have been evidenced through X-ray CT tomography. 

%Furthermore, we demonstrate that a gripper filled with a mixture of a small fraction of hard particles and a large amount of soft ones yields significantly large holding forces due to interlocking. Using X-ray imaging, we reveal the mechanism behind this observation: soft particles induce the gripper's volume reduction, displacing the stiffer particles around the object's protrusions, substantially increasing the gripper's resistance to bending and thus the force required to break the interlocking effect.

Our findings can advance the development of robotic grippers by addressing their greatest weakness: the hoisting force, which has limited their practical use until now.

%The outcomes of our study can be applied to reinforcing granular gripping systems, making them more reliable in manipulating diverse objects and rendering them resilient to changes in the target object geometry or surface properties.  

\begin{acknowledgements}
We gratefully acknowledge funding by Deutsche Forschungsgemeinschaft (DFG, German Research Foundation)--Project Number 411517575. The authors thank Walter Pucheanu for technical support.
\end{acknowledgements}

\section*{Compliance with ethical standards}
The authors declare that they have no conflict of interest.

\bibliographystyle{unsrtnat}

\bibliography{additionalBib.bib}   % name your BibTeX data base

\end{document}

%% file: color_define.tex
\definecolor{brickred}{rgb}{0.8, 0.25, 0.33}
\definecolor{darkorange}{rgb}{1.0, 0.55, 0.0}
\definecolor{persiangreen}{rgb}{0.0, 0.65, 0.58}
\definecolor{persianindigo}{rgb}{0.2, 0.07, 0.48}
\definecolor{cadet}{rgb}{0.33, 0.41, 0.47}
\definecolor{turquoisegreen}{rgb}{0.63, 0.84, 0.71}
\definecolor{sandybrown}{rgb}{0.96, 0.64, 0.38}
\definecolor{blueblue}{rgb}{0.0, 0.2, 0.6}
\definecolor{ballblue}{rgb}{0.13, 0.67, 0.8}
\definecolor{greengreen}{rgb}{0.0, 0.5, 0.0}

%% file: setup.tex
\begin{tikzpicture}
\small 
% \begin{scope}[local bounding box=redbox]
\node[] (setup)  at (0, 0) { \includegraphics[width=0.75\columnwidth]{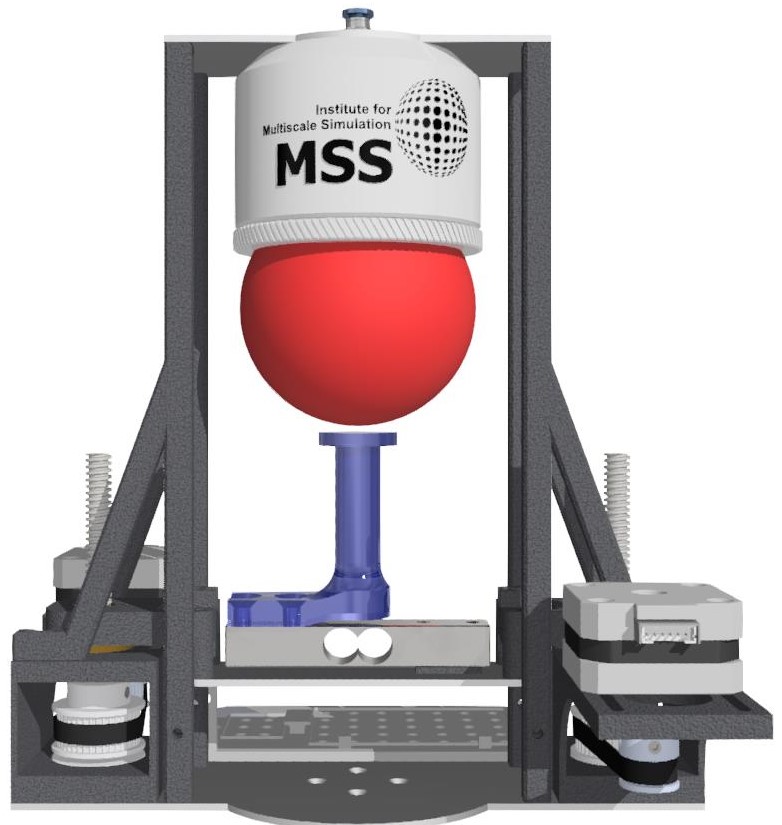}};

\node[right=-2mm of setup.34, anchor=west, text width=2.5cm, text=black] (holder) {gripper holder};
\node[left=26mm of holder] (holderPt) {};
% \draw[white, ultra thick](holder.west)--(holderPt);
\draw[-{Circle[fill=none]}, thick](holder.west)--(holderPt);

\node[right=-2mm of setup.15, anchor=west, text width=2.5cm, text=black] (gripper) 
{gripper filled \\with particles};
% \node[right=-67mm of setup.20, anchor=west] (holder2) {};
\node[left=28mm of gripper] (gripperPt) {};
% \draw[white, ultra thick](holder)--(holderPt);
\draw[-{Circle[fill=none]}, thick](gripper)--(gripperPt);

\node[right=-2mm of setup.-11, anchor=west, text width=2.5cm, text=black] (object) {object};
\node[left=33mm of object] (objectPt) {};
% \draw[white, ultra thick](holder.west)--(holderPt);
\draw[-{Circle[fill=none]}, thick](object.west)--(objectPt);

\node[right=-2mm of setup.-28, anchor=west, text width=2.5cm, text=black] (loadcell) {load cell};
\node[left=24mm of loadcell] (loadcellPt) {};
% \draw[white, ultra thick](holder.west)--(holderPt);
\draw[-{Circle[fill=none]}, thick](loadcell.west)--(loadcellPt);

%\node[right=-35mm of setup.20, anchor=west, text width=4.5cm, text=black] (holder) {\normalsize {membrane filled \\ with particles}};
%\node[left=16mm of holder] (holderPt) {};
%\draw[white, ultra thick](holder.west)--(holderPt);
%\draw[-{Circle[fill=white]}, ultra thick](holder.west)--(holderPt);

\node[right=-2mm of setup.-40, anchor=west,text width=2.5cm, text=black] (zstage) {platform};
\node[below=20mm of setup.center, xshift=-1mm] (zstagePt) {};
\node[left=21.5mm of zstage] (zstagePtmid) {};
\draw[-{Circle[fill=none]}, thick](zstage.west)--(zstagePtmid.center)--(zstagePt.center); 
%\draw[-{Circle[fill=none]}, thick](zstage.west)--(zstagePtmid); 

%\node[right=-2mm of setup.-40, anchor=west,text width=2.5cm, text=black] (zstage) {platform};
%\node[left=21.5mm of zstage] (zstagePtmid) {};
%\draw[-{Circle[fill=none]}, thick](zstage.west)--(zstagePtmid); 

% \end{scope}
% \draw[red] (redbox.south west)rectangle (redbox.north east);
\end{tikzpicture}
%\end{document}

%% file: holding-force-EPS.tex
\begin{tikzpicture}

\pgfplotsset{set layers}
\definecolor{mediumpersianblue}{rgb}{0.0, 0.4, 0.65}
\definecolor{pastelorange}{rgb}{1.0, 0.7, 0.28}
\definecolor{orangepeel}{rgb}{1.0, 0.62, 0.0}
%\pgfplotsset{every axis/.append style={font=\large,
%thick}}
    \begin{axis}[
      width=1.0\columnwidth, 
      height=0.30\textwidth,
      font=\normalsize,
%thick
    ybar=0pt, %xtick=data,
    ymin=0,ymax=19.2,
    xtick distance={2.5},
%    enlarge x limits = 0.1,
%    legend style={at={(0.5,-0.38)},
%    anchor=north,legend columns=-1},
%    ylabel style={align=center, text width=2cm},
    ylabel={holding force [\si{\newton}]},
    %symbolic x coords={wet-open, wet-close, dry-open, dry-close},
	xmin=0.0,xmax=6.4,
	%xtick={1,2,3,4},
    %xtick=\empty,
    %\node[] (setup)  at (0, 0)
	x tick style={draw=none},
    xtick={1.9, 4.5},
	xticklabels={ \small EPS, \small glass},
	%xticklabel style = {font=\footnotesize},
     %\node[label=above:{king} ] at (0.5,0.5) {0};
%    nodes near coords, 
	every node near coord/.style={
        opacity=1, 
        text depth=6.5mm,
        /pgf/number format/precision=2
        },
	every axis plot/.append style={
          ybar,
          bar width=1.3,
          bar shift=0pt,
          fill
        }
    ]

\addplot [fill = black, font=\scriptsize, opacity=0.75,
        error bars/.cd,
        y dir=both,
        y explicit relative, 
        error bar style={color=black, solid, opacity=1}
        ] 
coordinates {
    (1.9, 15.82) +- (0.15,0.15)
    };

\addplot [fill = black, font=\scriptsize, opacity=0.75,
        error bars/.cd,
        y dir=both,
        y explicit relative, 
        error bar style={color=black, solid, opacity=1}
        ] 
coordinates {
    (4.5, 4.54) +- (0.165,0.165)
	};
        
    \end{axis}

%\node at (4.9,1.9) {\includegraphics[width=1.1cm]{interlocking-object201.png}};

\end{tikzpicture}

%% file: softP-xray-interlocking.tex
\begin{tikzpicture}

\node[] (setup)  at (0, 0) {};

\node[] (setup2)  at (0, -4.2)
{\includegraphics[width=1.0\textwidth]{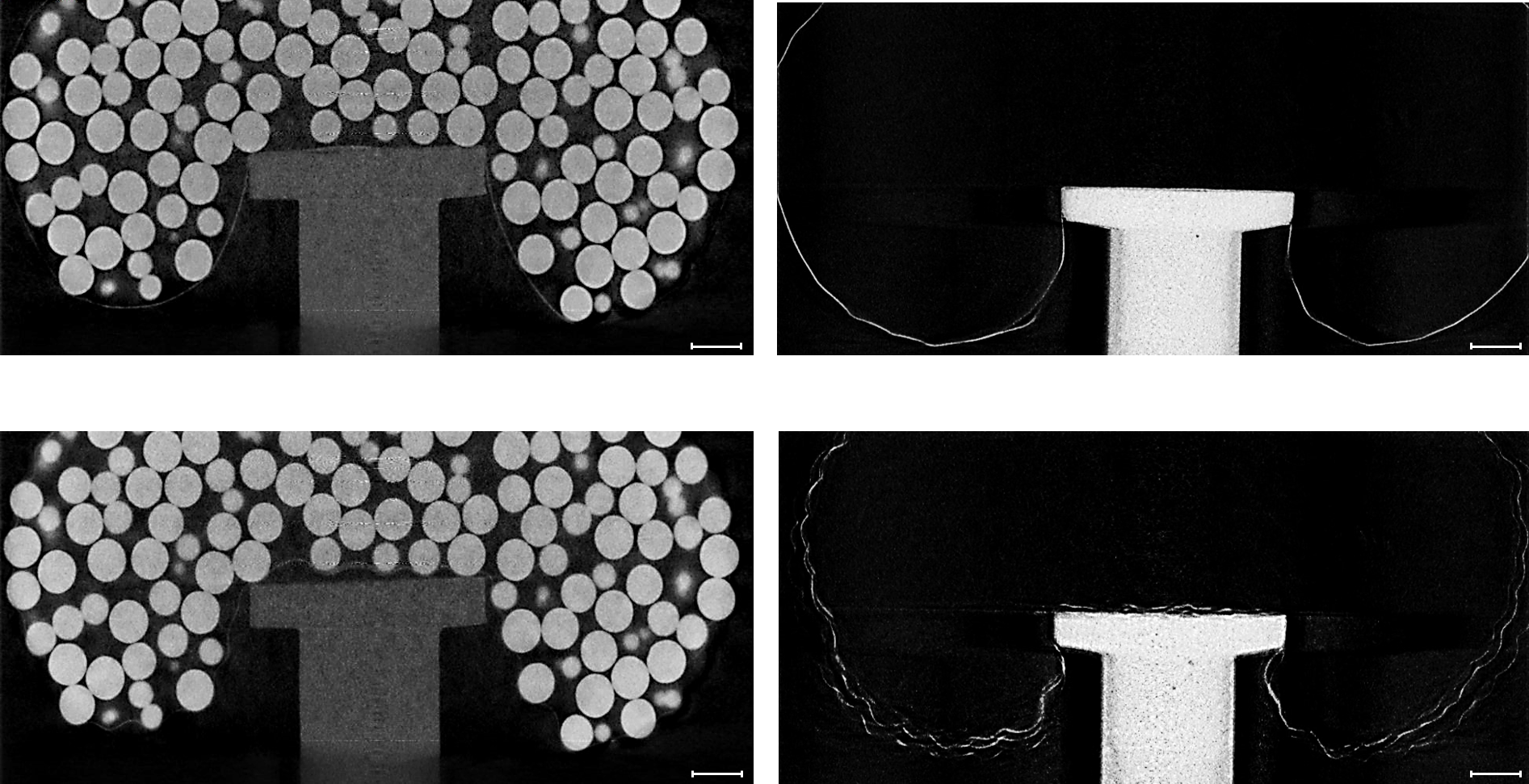}};

%\node[right=-27mm of setup.30, anchor=west, text width=6.5cm, yshift=5mm, text=black] (holder) {\large Before gripper evacuation};

%\node[right=-25mm of setup.-10, anchor=west, text width=6.5cm, yshift=-42mm, text=black] (holder) {\large After gripper evacuation};

\node[right=-89mm of setup.-50, anchor=west, text width=6.5cm, yshift=-34mm, text=white] (holder) {\large (a)};

\node[right=-10.5mm of setup.-50, anchor=west, text width=6.5cm, yshift=-34mm, text=white] (holder) {\scriptsize 5mm};

\node[right=-89mm of setup.-50, anchor=west, text width=6.5cm, yshift=-83mm, text=white] (holder) {\large (c)};

\node[right=-10.5mm of setup.-50, anchor=west, text width=6.5cm, yshift=-83mm, text=white] (holder) {\scriptsize 5mm};

\node[right=0mm of setup.-50, anchor=west, text width=6.5cm, yshift=-34mm, text=white] (holder) {\large (b)};

\node[right=78mm of setup.-50, anchor=west, text width=6.5cm, yshift=-34mm, text=white] (holder) {\scriptsize 5mm};

\node[right=0mm of setup.-50, anchor=west, text width=6.5cm, yshift=-83mm, text=white] (holder) {\large (d)};

\node[right=78mm of setup.-50, anchor=west, text width=6.5cm, yshift=-83mm, text=white] (holder) {\scriptsize 5mm};

\end{tikzpicture}

%% file: holding-force-mixtures2.tex
% used PGFPlots v1.16
\begin{tikzpicture}

\pgfplotsset{set layers}
\definecolor{mediumpersianblue}{rgb}{0.0, 0.4, 0.65}
\definecolor{pastelorange}{rgb}{1.0, 0.7, 0.28}
\definecolor{orangepeel}{rgb}{1.0, 0.62, 0.0}
%\pgfplotsset{every axis/.append style={font=\large,
%thick}}
    \begin{axis}[
      width=1.0\columnwidth, 
      height=0.30\textwidth,
      font=\normalsize,
%thick
    ybar=0pt, %xtick=data,
    ymin=0,ymax=34.2,
    xtick distance={2.5},
%    enlarge x limits = 0.1,
%    legend style={at={(0.5,-0.38)},
%    anchor=north,legend columns=-1},
%    ylabel style={align=center, text width=2cm},
    ylabel={hoisting force [\si{\newton}]},
    xlabel={fraction of glass beads},
    %symbolic x coords={wet-open, wet-close, dry-open, dry-close},
	xmin=0.0,xmax=34.0,
	%xtick={1,2,3,4},
    %xtick=\empty,
    %\node[] (setup)  at (0, 0)
	x tick style={draw=none},
    xtick={1.9, 5.4, 8.9, 12.8, 16.9, 20.9, 24.4, 27.9, 31.4},
	xticklabels={ \normalsize 0, \normalsize 0.05, \normalsize 0.1, \normalsize 0.15, \normalsize 0.20, \normalsize 0.25, \normalsize 0.5, \normalsize 0.75, \normalsize 1 },
	%xticklabel style = {font=\footnotesize},
     %\node[label=above:{king} ] at (0.5,0.5) {0};
%    nodes near coords, 
	every node near coord/.style={
        opacity=1, 
        text depth=6.5mm,
        /pgf/number format/precision=2
        },
	every axis plot/.append style={
          ybar,
          bar width=1.3,
          bar shift=0pt,
          fill
        }
    ]

\addplot [fill = blue, font=\scriptsize, opacity=0.75,
        error bars/.cd,
        y dir=both,
        y explicit relative, 
        error bar style={color=black, solid, opacity=1}
        ] 
coordinates {
    (1.9, 15.82) +- (0.15,0.15)
	};

\addplot [fill = black, font=\scriptsize, opacity=0.75,
        error bars/.cd,
        y dir=both,
        y explicit relative, 
        error bar style={color=black, solid, opacity=1}
        ] 
coordinates {
    (5.4, 18.72) +- (0.09,0.09)
    };

\addplot [fill = black, font=\scriptsize, opacity=0.75,
        error bars/.cd,
        y dir=both,
        y explicit relative, 
        error bar style={color=black, solid, opacity=1}
        ] 
coordinates {
    (8.9, 27.81) +- (0.14,0.14)
    };

\addplot [fill = black, font=\scriptsize, opacity=0.75,
        error bars/.cd,
        y dir=both,
        y explicit relative, 
        error bar style={color=black, solid, opacity=1}
        ] 
coordinates {
    (12.8, 18.13) +- (0.16,0.16)
	};

 \addplot [fill = black, font=\scriptsize, opacity=0.75,
        error bars/.cd,
        y dir=both,
        y explicit relative, 
        error bar style={color=black, solid, opacity=1}
        ] 
coordinates {
    (16.9, 17.14) +- (0.13,0.13)
	};

\addplot [fill = black, font=\scriptsize, opacity=0.75,
        error bars/.cd,
        y dir=both,
        y explicit relative, 
        error bar style={color=black, solid, opacity=1}
        ] 
coordinates {
    (20.9, 5.08) +- (0.22,0.22)
	};

\addplot [fill = black, font=\scriptsize, opacity=0.75,
        error bars/.cd,
        y dir=both,
        y explicit relative, 
        error bar style={color=black, solid, opacity=1}
        ] 
coordinates {
    (24.4, 4.69) +- (0.25,0.25)
	};

\addplot [fill = black, font=\scriptsize, opacity=0.75,
        error bars/.cd,
        y dir=both,
        y explicit relative, 
        error bar style={color=black, solid, opacity=1}
        ] 
coordinates {
    (27.9, 4.68) +- (0.16,0.16)
	};

\addplot [fill = blue, font=\scriptsize, opacity=0.75,
        error bars/.cd,
        y dir=both,
        y explicit relative, 
        error bar style={color=black, solid, opacity=1}
        ] 
coordinates {
    (31.4, 4.54) +- (0.165,0.165)
	};
        
    \end{axis}

%\node at (4.9,1.9) {\includegraphics[width=1.1cm]{interlocking-object201.png}};

\end{tikzpicture}

%% file: mixture-x-ray-interlocking.tex
\begin{tikzpicture}

\node[] (setup)  at (0, 0) {};

\node[] (setup2)  at (0, 0)
{\includegraphics[width=1.0\columnwidth]{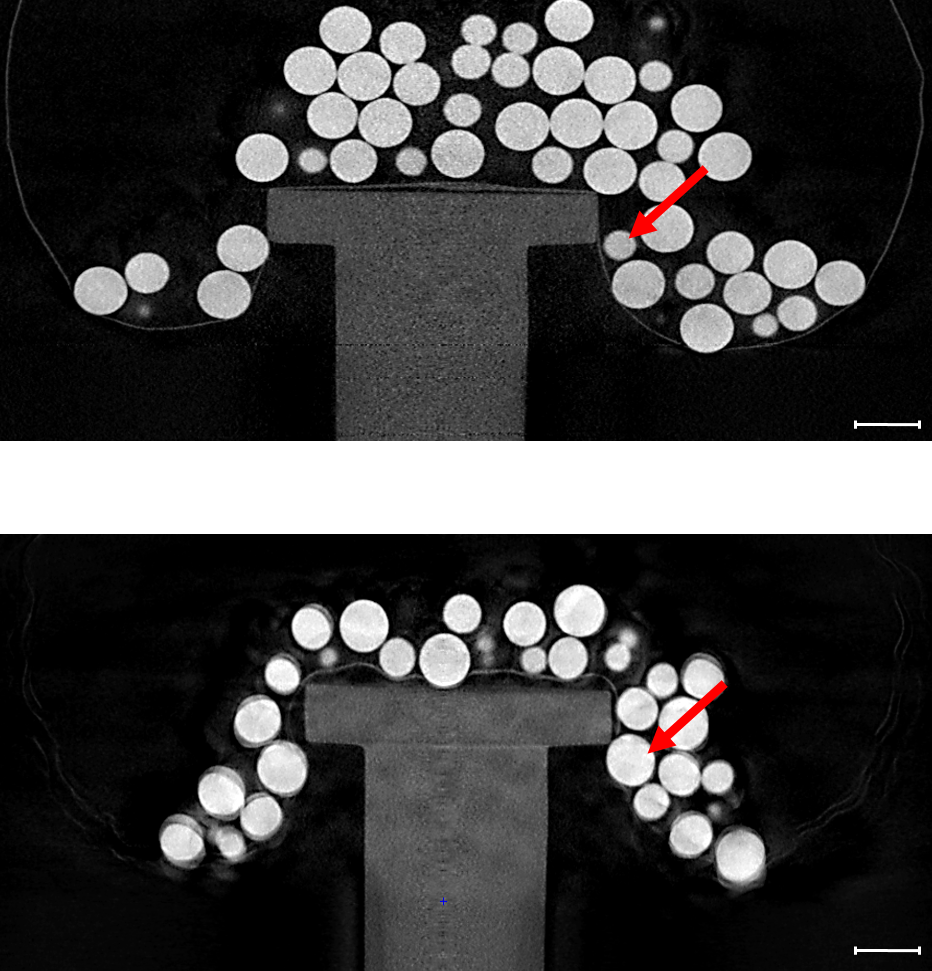}};

%\node[right=-23mm of setup.30, anchor=west, text width=6.5cm, yshift=53mm, text=black] (holder) {\large Before gripper evacuation};

%\node[right=-23mm of setup.-10, anchor=west, text width=6.5cm, yshift=0mm, text=black] (holder) {\large After gripper evacuation};

\node[right=-43mm of setup.-50, anchor=west, text width=6.5cm, yshift=9mm, text=white] (holder) {\large (a)};

\node[right=32.7mm of setup.-50, anchor=west, text width=6.5cm, yshift=-39mm, text=white] (holder) {\scriptsize 5mm};

\node[right=-43mm of setup.-50, anchor=west, text width=6.5cm, yshift=-39mm, text=white] (holder) {\large (b)};

\node[right=32.7mm of setup.-50, anchor=west, text width=6.5cm, yshift=8mm, text=white] (holder) {\scriptsize 5mm};

\end{tikzpicture}